\documentclass[amsmath,notitlepage,amssymb,aps,showkeys,floatfix,prd,a4paper,
  onecolumn,nofootinbib]{revtex4-2}

\usepackage{graphicx}
\usepackage{amsmath}
\usepackage{epstopdf}
\usepackage{amsfonts,amssymb}
\usepackage[utf8]{inputenc}
\usepackage{pstricks}
\usepackage{color}

\usepackage{array}

\newcolumntype{M}[1]{>{\centering\arraybackslash}m{#1}}
\newcolumntype{N}{@{}m{0pt}@{}}

\usepackage[compat=1.0.0]{tikz-feynman}
\usepackage{simplewick}
\usepackage{color, colortbl}
\definecolor{Gray}{gray}{0.9}
\usepackage{float}
\usepackage{tikz-feynman}
\usepackage{feynmp}
\usepackage{forest}

\usepackage{verbatim}    
\usepackage{dcolumn}
\usepackage{bm}
\usepackage{epsfig}
\usepackage{braket}
\usepackage[normalem]{ulem} 
\usepackage{longtable} 
\usepackage{multirow}

\newcommand{\be}{\begin{equation}}
\newcommand{\ee}{\end{equation}}
\newcommand{\ben}{\begin{eqnarray}}
\newcommand{\een}{\end{eqnarray}}
\newcommand{\lb}{\label}
\def\MeV{\mbox{ MeV}}

\def\MeV{\mbox{ MeV}}

\newcommand{\pslash}{\not{\hbox{\kern-2.3pt $p$}}}
\newcommand{\pdslash}{\not{\hbox{\kern-2pt $\partial$}}}

\begin{document}


\title{A note on the possible bound $D^{(\ast)} D ^{(\ast)}, \bar{B}^{(\ast)} \bar{B}^{(\ast)} $ and $D^{(\ast)} \bar{B}^{(\ast)} $  states}

\author{Luciano M. Abreu}
\email{luciano.abreu@ufba.br}
\affiliation{ Instituto de F\'isica, Universidade Federal da Bahia,
Campus Universit\'ario de Ondina, 40170-115, Bahia, Brazil}

\begin{abstract}
  
Motivated by the recently observation of the tetraquark $T_{cc}^+ $ state, in this work I revisit the Heavy-Meson Effective Theory to perform a simplified field-theoretical study of possible deuteron-like  $D^{(\ast)} D ^{(\ast)}, \bar{B}^{(\ast)} \bar{B}^{(\ast)} $ and $D^{(\ast)} \bar{B}^{(\ast)} $ molecules. In particular, using the data from $T_{cc}^+ $ as input to fix the potential associated to the shallow-bound $ D D^{*} $ system with $I(J^P) =  0(1^+)$, the conditions for the formation of other loosely-bound states in the doubly charmed sector are analyzed. The other sectors are also discussed. 
%

\end{abstract}

\maketitle

%
%
%
%


\section{Introduction}

Some months ago, the LHCb collaboration reported the detection of a new type of hadronic state containing two charm quarks, using data collected in proton-proton collisions, with statistical significance of more than $10 \,
\sigma$~\cite{LHCb:2021vvq,LHCb:2021auc}. It has been associated to  
a resonance seen in the $D^0 D^0 \pi^+ $-mass spectrum
with a mass of approximately $3875 \MeV$ and quantum numbers $J^P = 1^+$.
Its minimum valence quark content should be $c c \bar{u}\bar{d} $, giving it the status of the first observed doubly-charmed tetraquark. The data allows one to estimate the binding energy 
with respect to the $D^{*+} D^0 $ mass threshold and the decay width as $273\pm 61 \pm 5 _{-14}^{+11}$ keV and          
$410\pm 165 \pm 43 _{-38}^{+18}$ keV, respectively, in consonance with the expected properties of a $T_{cc}^+ $ isoscalar $J^P = 1^+$ tetraquark ground state
~\cite{Gelman:2002wf,Janc:2004qn,Vijande:2003ki,Navarra:2007yw,
Vijande:2007rf,Ebert:2007rn,Lee:2009rt,Yang:2009zzp,Hong:2018mpk,    
Hudspith:2020tdf,Cheng:2020wxa,Qin:2020zlg}. Since its detection, many works
appeared debating the possible mechanisms of its decay/formation and trying to
answer the question of its intrinsic nature~\cite{Agaev:2021vur,Dong:2021bvy, 
Agaev:2021vur,Dong:2021bvy,Feijoo:2021ppq,Huang:2021urd,Li:2021zbw,Ren:2021dsi,Xin:2021wcr, 
Yang:2021zhe,Meng:2021jnw,Ling:2021bir,Fleming:2021wmk,Jin:2021cxj, 
Azizi:2021aib,Hu:2021gdg,Abreu:2021jwm,Abreu:2022lfy,Dai:2021vgf,Albaladejo:2021vln,Du:2021zzh}. Due to the proximity of its mass to the $D^{*+} D^0 $ threshold, a natural interpretation that has 
been extensively used is to consider the $T_{cc}^+ $ as a deuteron-like molecule of $D^{*+} D^0 $. In this regard, it might be produced from the interactions of the charmed mesons and interpreted as a loosely-bound state, related to a pole just below the threshold and in the first Riemann sheet of the scattering amplitude.

In a historical perspective, the notion of molecular states with heavy-light 
hadrons was proposed more than four decades ago, in the analysis of interactions between the charmed and anti-charmed mesons \cite{Voloshin}. Afterwards, in subsequent decades this picture has been employed in different approaches, 
as in the quark-pion interaction framework for the investigations of several deuteron-like meson-meson bound states 
\cite{Tornqvist1,Tornqvist2}. Nevertheless, the discovery of exotic hadron states from 2003 onwards definitively stimulated the use of this concept in hadron physics. For instance, the most emblematic and intriguing exotic structure, 
the $X(3872) $, might have a non-negligible component associated to a shallow bound state of $D \bar{D}^{*}$
\cite{Tornqvist3,AlFiky,Braaten1,Dong1,Braaten2,Nieves,Hidalgo,
Guo,Alberto,XProd2,Braaten:2020iqw}; and so several other exotic states interpreted as hadron--anti-hadrons molecules have been proposed. We refer the reader to Refs.~\cite{Guo:2017jvc,Brambilla:2019esw} for reviews on this subject.

Additionally, prior to experimental evidence reported by LHCb mentioned above, possible hadron-hadron combinations to form doubly-charmed or doubly-bottomed bound states had also been speculated via use of different approaches, as one-boson-exchange models and frameworks based on heavy-quark flavor symmetry; see e.g.~\cite{Barnes:1999hs,Zhang:2007mu,Molina:2010tx,Vijande:2009kj,Yang:2009zzp,Du:2012wp,Li:2012ss,Luo:2017eub,Sakai:2017avl,Eichten:2017ffp,Xu:2017tsr,Wang:2018atz,Liu:2019zoy,Yu:2019sxx}. After this observation, this topic has naturally gained even more attention~\cite{Chen:2021cfl,Dai:2021vgf,Albaladejo:2021vln,Du:2021zzh,Zhao:2021cvg,Dai:2022ulk}. 

But there are places for other contributions on this issue. In the scenario of  heavy-meson chiral effective field theory, Refs.~\cite{Xu:2017tsr,Wang:2018atz} investigated the $D  D^* $ and $B B^* $ interactions at the next-to-leading order.  These interesting works have been performed before the $T_{cc}^+ $ observation, and their findings were not based on any experimental information, especially the $T_{cc}^+ $ mass and its small binding energy concerning the $D  D^* $ threshold in the molecular interpretation. Thus, this note has as the main purpose to perform a field-theoretical study of possible deuteron-like molecules  $D^{(\ast)} D ^{(\ast)}, \bar{B}^{(\ast)} \bar{B}^{(\ast)} $ and $D^{(\ast)} \bar{B}^{(\ast)} $ molecules, taking into account the data from $T_{cc}^+ $ as input to fix the potential associated to the shallow-bound $ D D^{*} $ system with $I(J^P) =  0(1^+)$. The conditions for the formation of other loosely-bound states in the doubly charmed sector are then analyzed. 
We estimate their masses and other properties, and compare them with 
those existing in literature.  A similar investigation is done in the sectors  $ \bar{B}^{(\ast)} \bar{B}^{(\ast)} $ and $D^{(\ast)} \bar{B}^{(\ast)} $, in analogy to the two-charmed sector.


\section{Formalism}
\lb{Formalism}
\subsection{Heavy-Meson Effective Theory}

As the interest here resides on the analysis of the possible loosely-bound states of $D^{(\ast)} D ^{(\ast)}, \bar{B}^{(\ast)} \bar{B}^{(\ast)} $ and $D^{(\ast)} \bar{B}^{(\ast)} $, we make use of an effective theory known as Heavy-Meson Effective Theory (HMET)
 \cite{AlFiky,Nieves,Manohar,Valderrama,Abreu:2015jma,Abreu:2016dfe,Abreu:2016xlr,Abreu:2017pos,Xu:2017tsr,Wang:2018atz}, which describes the interactions involving heavy mesons. 
We start by introducing the lowest order of the effective Lagrangian respecting heavy-quark spin, heavy-quark flavor and light-quark flavor symmetries,
\begin{eqnarray}
\mathcal{L} = \mathcal{L}_2 + \mathcal{L}_4;
\label{L1}  
\end{eqnarray} 
 the two-body piece is 
\begin{eqnarray}
\mathcal{L}_2 & = & - i \;\mathrm{Tr}\hspace{1pt} \left[ \bar{\mathcal{H}}  ^{(Q) a} v 
\cdot \partial \;\mathcal{H}_{a}  ^{(Q) } \right]
 - i \;\mathrm{Tr}\hspace{1pt} \left[\mathcal{H}  ^{(\bar{Q} ) a}
 v \cdot \partial  \; \bar{\mathcal{H}} _a ^{(\bar{Q} ) }
 \right] ,  \nonumber \\
\label{L2} 
\end{eqnarray} 
where $v$ denotes the velocity parameter; $\mathcal{H} _{a} ^{(Q/ \bar{Q})}$  the generalized superfields, defined as 
\begin{eqnarray}
\mathcal{H} _a ^{(Q)}= 
\left( \begin{array}{c}
H_a ^{(c)} \\
H_a ^{(b)} \end{array} \right);  \;\;\;\;\;
\mathcal{H}  ^{(\bar{Q}) a} =
\left( \begin{array}{cc}
H ^{(\bar{c}) a} , &
H ^{(\bar{b}) a} \end{array} \right), 
\label{H1}
\end{eqnarray}
with  $a$ being the doublet index of the isospin group  $SU(2)_V$; $Q = c,b$ the index with respect to the heavy-quark flavor group 
$SU(2)_{HF}$, and 
\begin{eqnarray}
H_a ^{(Q)} & = & \left( \frac{1+ v_{\mu} \gamma ^{\mu} }{2}\right)\left( 
P _{a \mu}^{* (Q)} 
\gamma ^{\mu}  - P _{a }^{ (Q)}  \gamma ^{5} \right) ,\nonumber \\
H ^{(\bar{Q}) a} & = & \left( P _{ \mu}^{* (\bar{Q}) a} \gamma ^{\mu} - 
P ^{ (\bar{Q}) a}  \gamma^{5} \right) \left( \frac{1 - v_{\mu} 
\gamma ^{\mu}}{2}\right).
\label{H2}
\end{eqnarray}
In the expression above, $P _{a }^{ (Q/ \bar{Q})}$ and $P _{a \mu}^{* (Q / \bar{Q})} $
are the pseudoscalar and vector heavy-meson fields forming a $\mathbf{\bar{2}}$
representation of $SU(2)_V$:
\begin{eqnarray}
  P _{a }^{ (c)}  =  \left( D^{0}, - D^{+} \right), \ \ \ \
  P _{a }^{ (\bar{c})} = \left( \bar{D}^{0},  D^{-} \right) ,
  \label{P1}
\end{eqnarray}
for the charmed mesons, and 
\begin{eqnarray}
  P _{a }^{ (b)}  =  \left( B^{-}, -\bar{B}^{0} \right), \ \ \ \
  P _{a }^{ (\bar{b})}  =  \left(  B^{+}, B^{0} \right) ,
  \label{P2}
\end{eqnarray}
for the bottomed mesons (and analogous expressions for the vector case).
To construct invariant quantities under the relevant symmetries, the hermitian conjugate fields $\mathcal{H}  ^{( Q / \bar{Q}) a}$ must be written as: 
\begin{eqnarray}
\bar{\mathcal{H}}  ^{(Q) a} & = & \gamma ^0 \mathcal{H}_a  ^{(Q) \dagger } \gamma ^0 
= \left( \begin{array}{cc}
\bar{H} ^{(c) a} ,
\bar{H} ^{(b) a} \end{array} \right) , \nonumber \\ 
\bar{\mathcal{H}}_a  ^{(\bar{Q}) }  & = & \gamma ^0 \mathcal{H}  ^{(Q) a \dagger }
 \gamma ^0
= \left( \begin{array}{c}
\bar{H}_a ^{(\bar{c})} \\
\bar{H}_a ^{(\bar{b})} \end{array} \right).
\label{H7}
\end{eqnarray}
The four-body interaction piece reads
\begin{eqnarray}
\mathcal{L}_4 & = & - \frac{D_1}{4} \mathrm{Tr}\hspace{1pt}
\left[ \bar{\mathcal{H}}  ^{(Q) a} 
 \mathcal{H}_{a} ^{(Q) } \gamma  ^{\mu} \right] \mathrm{Tr}\hspace{1pt} \left[
 \bar{\mathcal{H}}  ^{(Q) a} 
 \mathcal{H}_{a} ^{(Q) } \gamma  _{\mu} \right]  \nonumber \\
& & - \frac{D_2}{4} \mathrm{Tr}\hspace{1pt}\left[ \bar{\mathcal{H}}  ^{(Q) a} 
 \mathcal{H}_{a}  ^{(Q) } \gamma  ^{\mu} \gamma  ^{5} \right] \mathrm{Tr}\hspace{1pt} \left[
 \bar{\mathcal{H}}  ^{(Q) a} 
 \mathcal{H}_{a} ^{(Q) } \gamma  _{\mu} \gamma  ^{5}
  \right]   \nonumber \\
& &   - \frac{E_1}{4} \mathrm{Tr}\hspace{1pt} \left[ \bar{\mathcal{H}}  ^{(Q) a}
 (\tau)_{a}^{b}
 \mathcal{H} _{b} ^{(Q) } \gamma  ^{\mu} \right] 
 \nonumber \\
& & \times \mathrm{Tr}\hspace{1pt} 
    \left[
 \bar{\mathcal{H}}  ^{(Q) a}
  (\tau )_{a}^{b} \;\mathcal{H} _{b} ^{(Q) } \gamma  _{\mu} \right] 
   \nonumber \\
& &  - \frac{E_2}{4} \mathrm{Tr}\hspace{1pt} \left[ \bar{\mathcal{H}}  ^{(Q) a} 
 (\tau )_{a}^{b}
 \mathcal{H}_{b}  ^{(Q) } \gamma  ^{\mu} \gamma  ^{5} \right] \nonumber \\
& & \times  \mathrm{Tr}\hspace{1pt}
 \left[\bar{\mathcal{H}}  ^{(Q) a}  (\tau )_{a}^{b}
 \;\mathcal{H} _{b} ^{(Q) } \gamma  _{\mu} \gamma  ^{5}
  \right]  \nonumber \\
& & + \left[ \left( \bar{\mathcal{H}}  ^{(Q) a} , 
 \mathcal{H} _{a} ^{(Q) }  \right) \rightarrow  \left(  \mathcal{H}  ^{(\bar{Q})  a}, \bar{\mathcal{H}} _{a} ^{(\bar{Q})  }   \right) \right],  
  \label{L4}
  \end{eqnarray}
where $D_1, D_2, E_1, E_2$ are the low-energy constants (LECs) to be determined; $\tau$ are the Pauli matrices. The last line in Eq.~(\ref{L4}) represents the respective terms involving the heavy anti-quark fields necessary to guarantee the charge conjugation symmetry. Besides, I mention that the other Lorentz structures at leading order are not 
independent, and can be written as linear combination of the considered terms above. Thus, they will be omitted.

The heavy vector meson fields obey the transverse conditions: 
\begin{eqnarray}
 v \cdot  P _{a }^{* (Q)} & = & 0, \nonumber \\
  v \cdot  P ^{* (\bar{Q}) a} & = & 0 . 
\label{C1}  
 \end{eqnarray}
They define the three different polarizations of the heavy vector mesons.

Since the present study relies on the leading order in the $1/m_Q$ expansion,  
relativistic effects are suppressed. As a consequence, the non-relativistic version of the theory can properly describe two-heavy meson systems. Accordingly, for convenience the velocity parameter chosen is $v = \left( 1, \vec{0} \right)$, while the normalization adopted is~\cite{Manohar,Valderrama}:
\be 
  \sqrt{2}  P _{a }^{(* \mu)} \rightarrow  P _{a }^{(* \mu) }, 
\label{norm}
\ee
Noticing that the component $P _{a }^{(* 0)}$ becomes irrelevant, only the  
the Euclidean part of the vector meson fields will be taken into account henceforth. 

\subsection{Transition Amplitudes}

In order to obtain the amplitudes for the processes 
\be 
P ^{(\ast)(Q)} (1)  P ^{(\ast)(Q)} (2) \rightarrow P ^{(\ast)(Q)}(3) P ^{(\ast)(Q)} (4), 
\label{scat}
\ee
we make use of the Breit approximation to relate the non-relativistic interaction potential, $V$, and the scattering amplitude $ i \mathcal{M}
(P ^{(\ast)}P ^{(\ast)} \rightarrow P ^{(\ast)} P ^{(\ast)})$:
\ben 
V(\vec{p}) = - \frac{1}{\sqrt{\Pi _i 2 m_i \Pi _f 2 m_f}} \mathcal{M}(P ^{(\ast)}P ^{(\ast)} \rightarrow P ^{(\ast)} P ^{(\ast)}), \nonumber \\
\label{rel1} 
\een
where $m_i $ and $m_f$ are the masses of initial and final states, and $\vec{p}$
is the momentum exchanged between the particles in Center-of-Mass frame.

Focusing on the description of possible $S$-wave bound states, following Refs.~\cite{AlFiky,Nieves,Guo,Valderrama,Abreu:2015jma} I assume that at tree level the pion-exchange effects are not relevant. This can be justified as follows:  at lowest order the integration of the pion degrees of freedom might be interpreted as a modification in the LEC $D_2$ in Eq.~(\ref{L4}) by a factor proportional to $ (g_{\pi}/f_{\pi})^2$ ($ g_{\pi} $ being the bare coupling constant of the $ P ^{\ast}P  \pi $ and $ P ^{\ast} P ^{\ast}  \pi $ vertices, and $ f_{\pi} $ the bare pion decay constant). In this sense, the matching condition for the LECs necessary to yield a given bound state can provide the modified $D_2$ already carrying the mentioned contribution from pion integration.
Then, we can explore the leading-order potential of HMET only with contact interactions present in Eq. (\ref{L4}), and investigate the region of parameter space where the pion-exchange contribution is irrelevant. The consequence of disregarding the
details of the interaction is that the present approach is restricted to the context of shallow bound states and bigger scattering lengths. This fact, however, does not impose any severe problem, keeping in mind that the point here is to explore the formation of possible loosely bound states with small binding energies, analogously to the molecular interpretation of the recently observed doubly charmed tetraquark $T_{cc}^+ $ state. Calculations of higher-order pionic corrections to the matching conditions, which are beyond the scope of this study, can be used to investigate possible states with higher binding energies.

Thus, it is possible to obtain  
the scattering amplitude at tree-level approximation from the four-body Lagrangian $\mathcal{L}_4$ in Eq. (\ref{L4}), 
yielding the effective potential $V $ in the basis of states 
$\mathcal{B}
\equiv \left\{  | P P  \rangle,
 \, | P^{*} P \rangle, \, | P P^{*}\rangle, \, 
| P^{*} P^{*} \rangle  \right\} $.
 It is shown in Table 
\ref{table1}.

\begin{table}[htp]
  \caption{Terms of interaction potential $V (\vec{p})$ in the basis 
$\mathcal{B} 
\equiv \left\{  | P P  \rangle, \,
 | P^{*} P \rangle, \, | P P^{*}\rangle, \, 
| P^{*} P^{*} \rangle  \right\} $ (we have omitted the heavy-quark flavor index for convenience). $\vec{\varepsilon}_i$ means the polarization 
of incoming or outgoing vector heavy meson; $\vec{S}_i$ is the spin-1 operator, 
whose matrix elements are equivalent to the vector product of polarizations, i.e.: $ \vec{S}_1 \equiv   (\vec{\varepsilon}_3 ^{*}  \times \vec{\varepsilon}_1) $ and $ \vec{S}_2 \equiv   (\vec{\varepsilon}_4 ^{*}  \times \vec{\varepsilon}_2) $; and $C_i  \equiv  D_i + E_i \ \tau \cdot \tau$~\cite{Valderrama,Abreu:2015jma}.  
}
\begin{center}
\begin{tabular}{c|c|c|c|c} \hline \hline
 & $P_1 P_2 $ & $P^{*} P$ &  $P P^{*}$  &  $ P^{*} P^{*}$ \\   \hline
$P P $ & $C_1$  &  0    &    0   & 
$- C_2 \vec{\varepsilon}_1 \cdot \vec{\varepsilon}_2 $\\
$P^{*} P$ & 0  & $C_1 \vec{\varepsilon}_3 ^{*} \cdot \vec{\varepsilon}_1$ 
   & $- C_2 \vec{\varepsilon}_3 ^{*} \cdot \vec{\varepsilon}_2$
& $- C_2 \vec{\varepsilon}_2 \cdot \vec{S}_1$ \\
$ P  P^{*} $ & 0  &  $- C_2 \vec{\varepsilon}_4 ^{*} \cdot \vec{\varepsilon}_1 $  
 & $ C_1 \vec{\varepsilon}_4 ^{*} \cdot \vec{\varepsilon}_2 $
 & $ C_2 \vec{\varepsilon}_1 \cdot \vec{S}_2$ \\
 $P^{*} P ^{*}$ & $- C_2 \vec{\varepsilon}_3  ^{*}\cdot \vec{\varepsilon}_4 ^{*}$ 
&  $- C_2 \vec{\varepsilon}_4 ^{*} \cdot \vec{S}_1$
& $C_2 \vec{\varepsilon}_3 ^{*} \cdot \vec{S}_2$ 
& \begin{tabular}{@{}c@{}}$C_1 \vec{\varepsilon}_3 ^{*} \cdot \vec{\varepsilon}_1
\vec{\varepsilon}_4 ^{*} \cdot \vec{\varepsilon}_2 $ 
\\ $+ C_2\vec{S}_1 \cdot 
\vec{S}_2$\end{tabular}
   \\
 \hline \hline
\end{tabular}
\end{center}
\label{table1}
\end{table}

Concerning the relation between the isospin and particle bases, 
naively one can consider that each state in the $\mathcal{B} $-basis is composed of the isosinglet $ |[P ^{(\ast)(Q_1)} P ^{(\ast)(Q_2)}]^{ I=0} \rangle $, 
\ben 
\frac{1}{\sqrt{2}} 
\left[ | P_1^{(\ast)(Q_1)} P_2 ^{(\ast)(Q_2)} 
+ | P_2 ^{(\ast)(Q_1)} P_1 ^{(\ast)(Q_2)}  \rangle \right] ;
\label{iso}
\een
and the isotriplet $ | [P ^{(\ast)(Q_1)} P ^{(\ast)(Q_2)}]^{ I=1}  \rangle  $, 
\ben 
& & \lbrace 
| P_1 ^{(\ast)(Q_1)} P_1 ^{(\ast)(Q_2)} \rangle , 
 \frac{1}{\sqrt{2}}  \left[ | P_1 ^{(\ast)(Q_1)} P_2 ^{(\ast)(Q_2)} \rangle 
\right. \nonumber \\ 
& & - \left.
| P_2 ^{(\ast)(Q_1)} P_1 ^{(\ast)(Q_2)}  \rangle \right] , 
|  P_2 ^{(\ast)(Q_1)} P_2 ^{(\ast)(Q_2)} \rangle \rbrace, 
\label{tri}
\een  
where $ P_1 ^{(\ast)(Q)}, P_2 ^{(\ast)(Q)} $ are the components of the doublets in Eqs.~(\ref{P1}) and (\ref{P2}).  This is valid for the di-meson states with both open charm and bottom $D^{(\ast)} \bar{B}^{(\ast)} $ .
However, identical di-meson structures (i.e. two charmed [or bottomed] mesons $ D^{(*)} D^{(*)} \ [ \bar{B}^{(*)} \bar{B}^{(*)}]$) must be considered separately. The total wave function for the systems $ D D, D^{*} D^{*} \ [ \bar{B}  \bar{B},\bar{B}^{*} \bar{B}^{*} ]$ should be symmetric under the combination of rotation, isospin and heavy-quark spin groups~\cite{Ke:2021rxd,Chen:2021cfl}.
This engenders a selection rule given by $L + S + I + 1 = \textrm{even number},$
where $L$ is the orbital angular momentum between the mesons 
(for the $ S $-wave $L=0$); $ S $ and $ I $ are the total spin and isospin, respectively. In the end, the allowed $ S $-wave doubly-charmed meson systems are, using the notation $ |[D^{(*)} D^{(*)}]_{S}^{I}\rangle $: 
\ben
& & |[D D]_{0}^{1}\rangle, | [D D^{*}] _{1}^{0,1}\rangle,
| [D^{*} D^{*}]_{0,2}^{1}\rangle, | [D^{*} D^{*})]_{1}^{0} \rangle , 
\label{allstates}
\een
and similarly for $ S $-wave doubly-bottomed meson structures.

The coupling constants $C_1$ and $C_2$ in Table 
\ref{table1} are given by the following 
expressions with respect to the specific channels of isospin $SU(2)_V$ basis:
\ben 
|[P ^{(\ast)(Q_1)} P ^{(\ast)(Q_2)}]_S ^{0} \rangle & : &  C_i = 2 D_i - 6 E_i, \nonumber \\
|[P ^{(\ast)(Q_1)} P ^{(\ast)(Q_2)}]_S ^{1} \rangle & : &  C_i = 2 D_i + 2 E_i .
\label{C2} 
\een

To check the formation of possible dynamically generated poles, one should analyze the solutions of the Lippmann-Schwinger equation
\ben
T ^{(\alpha \beta)} = V ^{(\alpha \beta)} + \int \frac{d^{4}q}{(2 \pi )^{4}} 
V^{(\alpha \gamma)} \,G \, T ^{(\gamma \beta)}, 
\label{LS1}
\een
where $\alpha, \beta, \gamma = |[P ^{(\ast)(Q_1)} P ^{(\ast)(Q_2)}]_S ^{I} \rangle $
 represent each channel associated to the 
 $ \mathcal{B}$  and
 isospin  bases, respectively; $ G $ is the loop function in the non-relativistic context,
 given by
\ben 
G & \equiv & 
\frac{1}{\frac{\vec{p}^{2}}{2 m_{P_1 ^{(\ast)}} } + 
q_0 - \frac{\vec{q}^{2}}{2 m_{P_1 ^{(\ast)}} } + i \epsilon} \;\;
 \nonumber \\ 
& &  \times
\frac{1}{\frac{\vec{p}^{2}}{2 m_{P_2 ^{(\ast)}} } + 
q_0 - \frac{\vec{q}^{2}}{2 m_{P_2 ^{(\ast)}}}  + i \epsilon }. 
\label{G1}
\een
For vector mesons, we must perform the replacement
  $ G \rightarrow G^{\mu \nu } $. For a specific channel, the Lippmann-Schwinger equation can be written as
\ben 
T ^{ (\alpha) } = \frac{V ^{ (\alpha) } }{1 - V ^{ (\alpha) } G ^{ (\alpha) } }. 
\label{LS2}
\een
Its pole structure is characterized as follows: resonances are 
understood as the poles located in the fourth quadrant of the momentum complex plane (in the second Riemann sheet), while bound states are below the threshold (in the first Riemann sheet). Centering on the bound-state solutions, and employing  the residue theorem and dimensional regularization, after some manipulations Eq.~(\ref{LS2}) reads~\cite{AlFiky} 
\ben 
T ^{ (\alpha) } = \frac{\tilde{V} ^{ (\alpha) } }{1 + \frac{i}{8 \pi} 
\mu |\vec{p}| \;\tilde{V} ^{ (\alpha) } } ;
\label{CS3}
\een
where $\tilde{V} ^{ (\alpha) }$ is the renormalized potential, 
and  $\mu  $ is reduced mass of the di-meson system. Obviously the $\tilde{V} ^{ (\alpha) }$  depends on the renormalization scheme chosen, but the observables to be introduced below should be renormalization-independent. So, here the dimensional regularization is used for convenience. 
To simplify the notation, we continue to denote the 
renormalized potential as $\tilde{V} ^{ (\alpha) } \rightarrow V ^{ (\alpha) }$, and accordingly for the renormalized couplings $D_1, D_2, E_1 $ and $ E_2$.

From Eq. (\ref{CS3}) the position of the pole $E_{Pole}$ on the energy scale can be inferred, and noticing that in the present case the energy is measured with respect to the mass $m_{P_1 ^{(\ast)}} + m_{P_2 ^{(\ast)}}$, then the mass of the bound state is $M = m_{P_1 ^{(\ast)}} + m_{P_2 ^{(\ast)}} - E_{Pole}$. Hence, $ E_{Pole} $ can be interpreted as the binding 
energy, being explicitly given by
\ben 
E_b ^{ (\alpha) } = \frac{32 \pi^{2} }{\left( V ^{ (\alpha) } \right)^2 
\mu ^3 }, 
\label{BE1}
\een
The scattering length can be derived as well,  
\ben 
a_s ^{ (\alpha) } \equiv \left( 2 \mu E_b ^{ (\alpha) } \right)^{-\frac{1}{2}} = \frac{ \mu V ^{ (\alpha)  }  }
{ 8 \pi } .
\label{SL}
\een

\section{Results}
\lb{Results}

With the treatment for the Lipmann-Schwinger equation outlined  above, the possibility of formation shallow $S$-wave bound states of di-mesons can be concretely investigated. It is worth emphasizing that the study is restricted to the parameter region of 
relevance of contact-range interaction, where the one pion-exchange (OPE) 
contribution is not relevant. 
To better understand this limitation, I benefit from the analyses of Refs~\cite{Valderrama,Fleming:2007rp,Meng:2022ozq}. According to them, the leading-order contributions coming from OPE for the channels $ 
P ^{\ast}P \rightarrow P P ^{\ast}$ and $ 
P ^{\ast} P^{\ast} \rightarrow P^{\ast} P ^{\ast}$ are proportional to the propagator of the pion as follows, 
\ben 
\left[ p^2 - m_{\pi}^2 \right] ^{-1} \approx - \left[ \vec{p}^2 + \mu_{\pi}^2 \right] ^{-1}
\label{pion_prop}
\een
where $ p $ is the exchanged
momentum between the heavy mesons, and $ \mu_{\pi}  $  the effective pion mass to be
used in each case. We observe that the channel $ 
P P \rightarrow P P$ does not have OPE contributions due to the lack of a vertex involving three pseudoscalar mesons. Then, the range of this pion-less effective theory is $ |\vec{p}|  < \mu_{\pi} $. 
 For the channel $ P ^{\ast} P^{\ast} \rightarrow P^{\ast} P ^{\ast} $, we have  $ \mu_{\pi} = m _{\pi}$ ($ m_{\pi} $ being the physical pion mass). However, for the channel  $ P ^{\ast} P \rightarrow P P ^{\ast}$, the squared effective pion mass  in the static approximation becomes  $ \mu_{\pi}^2 \equiv m_{\pi}^2 - ( m_{ P^*} - m_{ P} )^2  $, since in the propagator is neglected the energy dependence other than the mass splitting of $ P^* $ and $ P $ (i.e. $p^0 \approx m_{ P^*} - m_{ P}$).
In the charm sector,  $ \mu_{\pi} $ is anomalously smaller ($ \mu_{\pi}  < m_{\pi} $)\footnote{As in Ref.~\cite{Valderrama}, if the interest holds only on the wave functions, the imaginary piece of the OPE potential for $  \mu_{\pi} ^2 < 0 $ engendered in charm sector can be ignored and the static limit remains useful.}; this is not the case of the bottom sector: $ \mu_{\pi} \approx m_{\pi} $. 
Additionally, it can be noticed that in the heavy quark limit $ m_Q \rightarrow \infty $ ($ m_Q $ being the heavy quark mass), one gets the expansion of the potential in powers of $ 1/m_Q $, since $ \mu_{\pi} \approx m_{\pi}  + O(1/m_Q) $~\cite{Valderrama}. Thus, although in the charm sector the $ P ^{\ast}P \rightarrow P P ^{\ast}$ potential in a pion-less treatment requires a smaller effective pion mass, for the purposes of this study the $ m_{\pi} $ stands as a useful scale in the context of static approximation and heavy quark limit, keeping in mind that the interest here is in systems with small binding energy, which from the discussion done in previous section generates large length scales, i.e. $ a_s ^{ (\alpha) } >  \lambda _C \equiv 1/m_{\pi}$ ($ \lambda _C  $ being pion Compton wavelength). In other words, the smaller binding energy, the larger the scattering length, and therefore the requirement $ |\vec{p}|  < \mu_{\pi} $ is easily fulfilled and long-range effects yielded by pions might be neglected.

In this sense, the possible $S$-wave bound structures will be characterized under the evaluation of relevant observables and their dependence with the parameters in each sector.
In this Section will be used the isospin-averaged masses for the heavy mesons reported in Ref.~\cite{ParticleDataGroup:2020ssz}.

\subsection{ $ |[D^{(*)} D^{(*)}]_{S}^{I}\rangle $ systems}

Beginning with the doubly-charmed systems, it can be seen in Table \ref{table1} that the transition amplitudes depend on the constants  
$C_1$ and $C_2$ , i.e. on $D_1, E_1, D_2  $ and $E_2$, according to the corresponding 
 channel of effective potential. First, a general discussion on the parameter space can be performed. Due to its richness, we restrict ourselves to explore it by taking some assumptions, namely: the first one is that the bound states should have binding energy greater 
than 0.1 MeV. This is justified by the fact that  $ E_b $ should have a sufficient magnitude to distinguish the masses of bound states from the respective thresholds. Smaller values $ E_b $ imposes some difficulties in the interpretation of loosely-bound states. Taking for example the case of $T_{cc}^+ $, the experimental error reported by the LHCb is $ | E_b^{(error)} | \lesssim 0.1 $ MeV, since $E_b ^{(Exp)}= 273\pm 61 \pm 5 _{-14}^{+11}$ keV. Thus, it seems reasonable to analyze the parameter space with the considered constraint.   The second one is the condition $ a_s ^{ (\alpha) } >  \lambda _C $ (as discussed before). 
In this region the relevant parameters acquire values that allow loosely bound systems 
for the isospin states explicited in Eq. (\ref{C2}). 

\begin{figure}[!ht]
	\centering
           \includegraphics[{width=8.0cm}]{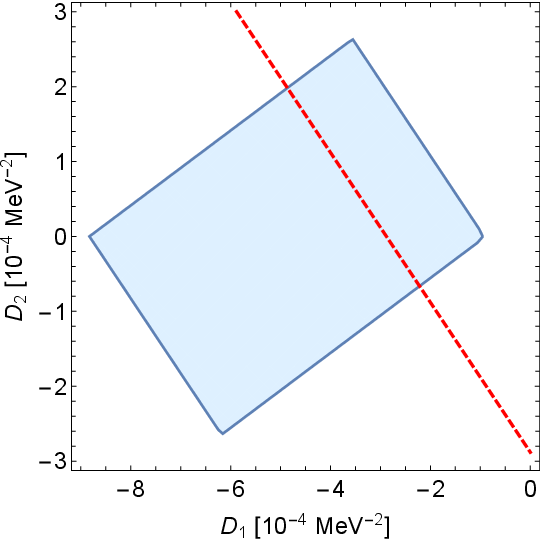}
\caption{$(D_1,D_2)$-parameter 
space (with $E_1, E_2 = 0$); light shaded area represent the region in which the parameters acquire 
values that allow shallow $S$-wave bound states for the $ |[D^{(*)} D^{(*)}]_{S}^{I}\rangle $ systems given in Eq.~(\ref{allstates}),  with binding energy greater 
than 0.1 MeV and obeying the condition $ a_s ^{ (\alpha) } >  \lambda _C $ . The dashed line represents the relation between the parameters which reproduces the values used for the system $ | [D D^{*}] _{1}^{0}\rangle $, interpreted as the $T_{cc}^+ $ state. }
	\label{FIGX1}
\end{figure}

As a first attempt, the analysis will be limited to the $(D_1,D_2)$-parameter space, by taking vanishing values for $E_1, E_2 $. It is  show in Fig.~\ref{FIGX1} the light shaded area indicating the intersection region  in which the parameters acquire values that allow  shallow $S$-wave bound states for the systems given in Eq.~(\ref{allstates}), in consonance with the above mentioned requirements.

Second, let us look at the experimental side. Here it appears a severe difficulty to investigate the parameter space in more detail, since the available experimental data in doubly-heavy quark systems are very scarce.
 Specifically, at the best of my knowledge there is only the very recently observed $T_{cc}^+ $ state. As pointed in Introduction, its expected quantum numbers are $I(J^P) =  0(1^+)$, and its mass 
is $273\pm 61 \pm 5 _{-14}^{+11}$ keV below the threshold of $D^{*+} D^0 $~\cite{LHCb:2021vvq,LHCb:2021auc}. Therefore, a natural interpretation for the $T_{cc}^+ $ is to consider it as a hadron molecule of $D^{*} D $. Accordingly, this information can be used as input to fix the potential associated to the $ | [D D^{*}] _{1}^{0}\rangle $.  In the calculations 
 we consider the following central values of $T_{cc}^+ $ mass, threshold and binding 
 energy: $M_X = 3874.75$ MeV, $ m_{D} + m_{\bar{D}^*} = 3875.8$ MeV  \cite{ParticleDataGroup:2020ssz}, and $E_b = 1.05$ MeV, respectively (stressing that the isospin-averaged masses have been used for the charmed mesons). Using this information in Eqs. (\ref{BE1}) and (\ref{SL}), we obtain the relation $V^{[D D^{*}] _{1}^{0}} = -2 D_1 + 6 E_1 - 2 D_2 + 6 E_2 = 5.76 \times 10^{-4} \MeV ^{-2}$. Taking $E_1, E_2 = 0$, in $(D_1,D_2)$-parameter space this relation is denoted in Fig.~\ref{FIGX1} by the dashed red line. It can be noticed that since bound-state solutions require $V^{[D D^{*}] _{1}^{0}} >0 $, non-vanishing values for $E_1, E_2 $ perform a  displacement of the dashed line; 
the increasing (decreasing) of the magnitude of $E_1,E_2$ induces 
a decreasing (increase) of values of $D_1,D_2$ parameters to get bound states. 

From the formalism and assumptions described above, the mass, binding energy and scattering length of the 
$ |[D^{(*)} D^{(*)}]_{S}^{I}\rangle $ systems can be estimated by matching the parameters to reproduce the $ | [D D^{*}] _{1}^{0}\rangle $ bound structure. Then, as an example in Table~\ref{table2} is shown the results obtained with three sets of parameters in a such way that the first two ones have their magnitudes along the dashed line inside the  region of intersection in Fig.~\ref{FIGX1}, whereas the latter one has its magnitude along the line but outside this area. The findings suggest that all the bound states are just near the thresholds of the related channels, but this proximity depends on the set of parameters. The set $ (ii) $ takes $D_2,E_2 = 0$, similarly to the estimate done in Ref.~\cite{Xu:2017tsr} in the analysis of the finite-range potential $V(r)$, and engenders higher binding energies and smaller scattering lengths. In contrast, the set $ (iii) $ admits only the bound state solution for the system $ | [D D^{*}] _{1}^{0}\rangle $. 

\begin{center}
\begin{table}[h!]
 \caption{ Relevant quantities for the $ |[D^{(*)} D^{(*)}]_{S}^{I}\rangle $ systems. $M_{Th}$, $M$, $ E_b $, and $a_s$ mean threshold mass, pole position,  binding energy and  scattering length of the respective state. The results are obtained in the context of three different set of parameters $(D_1,E_1,D_2,E_2)$ (values in $ \MeV ^{-2}$): $(i) (-0.000238,0,-0.00005,0);$ $ (ii) (-0.0006,-0.0001,0,0) ;$ and $ (iii) (-0.001,0.001,0.0067,  0.001) $. All units  in the Table are in MeV, except in the case of the scattering length, which is in fm. The isospin-averaged masses for the heavy mesons reported in Ref.~\cite{ParticleDataGroup:2020ssz} have been used in the calculations. Bold row denotes the state interpreted as the tetraquark $T_{cc}^+ $. The symbol ``-" indicates no bound state due to the non-existence of bound solution or the disobedience of the requirements mentioned in the text. } 
\begin{tabular}{M{60pt}|M{35pt}|M{35pt}|M{35pt}|M{35pt}}
\hline\hline \vspace{3pt}
System    & Set & $ M $ &  $ E_b $  & $a_s$  \\[3pt]
   ($M_{Th}$)  &  & &   &  \\[3pt]
\hline 
\hline \vspace{3pt}
$ |[D D]_{0}^{1}\rangle  $  & (i) & 3732.94 &  1.54 &  3.61  \\[2pt]
 (3734.48) &   (ii)  & 3734.28  & 0.20  &  10.30
 \\[2pt]
  &   (iii)  & -  & -  &  -
 \\[2pt]\hline \vspace{3pt}
$\mathbf{ |[D D^*]_{1}^{0}\rangle  }$  & \textbf{(i)} & \textbf{3874.75} & \textbf{1.05} & \textbf{4.37}  \\[3pt]
\textbf{(3875.80)}  &   \textbf{(ii)} & \textbf{3874.75}  & \textbf{1.05}  & \textbf{4.37} 
 \\[2pt]
  &    \textbf{(iii)}    & \textbf{3874.75}  & \textbf{1.05}  & \textbf{4.37} 
 \\[2pt] \hline \vspace{3pt}
 $ |[D D^*]_{1}^{1}\rangle  $  & (i) & 3873.34 &  2.46  & 2.85   \\[2pt]
(3875.80)  &   (ii) & 3875.62 &  0.18  & 10.68  
 \\[2pt] 
 &   (iii)  & -  & -  &  -
 \\[2pt] \hline \vspace{3pt}
$ |[D^* D^*]_{0}^{1}\rangle  $  & (i) & 4013.03 & 4.09 &  2.17 \\[2pt]
(4017.12)  &   (ii)  & 4016.96  & 0.16  & 11.08 
 \\[2pt]
  &   (iii)  & -  & -  &  -
 \\[2pt]\hline \vspace{3pt}
$ |[D^* D^*]_{1}^{0}\rangle  $  &  (i) & 4014.92 &  2.20 & 2.96  \\[2pt]
(4017.12)   &   (ii)  & 4016.18  &  0.94 & 4.53  
\\[2pt]&   (iii)  & -  & -  &  -
 \\[2pt]
 \hline \vspace{3pt}
$ |[D^* D^*]_{2}^{1}\rangle  $    &   (i)  &  4016.18 & 0.94 & 4.54 \\[2pt]
 (4017.12) & (ii)  & 4016.96  & 0.16  & 11.08 
  \\[3pt]
 &   (iii)  & -  & -  &  -
 \\[2pt]\hline 
\hline \end{tabular}
  \label{table2}
\end{table}
\end{center}

A brief note must be consecrated to the errors involved in this estimation. The relevant quantities are obtained by taking into account the violation of heavy quark spin symmetry, 
due to the finite heavy quark masses. This 
effect produces the most relevant high-order corrections to the heavy meson contact 
interactions. We expect a relative uncertainty of the order of $\Lambda _{QCD}/m_Q$ 
in the values of $\tilde{V} ^{ (\alpha) }$ in heavy-quark limit. So, taking 
$\Lambda _{QCD} \sim 200 $ MeV and the quark charm mass $m_c \sim 1.5$ GeV, this estimated error is of 15\% in leading order contact 
interactions. This gives in principle non-negligible uncertainties in the quantities given in Table~\ref{table2}. Notwithstanding, since the purpose here is not to make very accurate predictions, the errors in the results have been omitted.

For the sake of performing a comparison with existing literature, let us now look at the studies previous to the detection of $T_{cc}^+ $. Coming back to the Ref.~\cite{Xu:2017tsr},  an interesting analysis of $D  D^* $ interactions with pion-exchange contributions has been developed by using heavy meson spin symmetry relations. Due the lack of experimental information at that time, the cited work determined the LECs for the contact contributions through the resonance saturation model, in which has been  assumed that these short-range couplings result from the light (scalar, vector and axial) meson exchanges. As a consequence, the central values for the LECs have been: $(D_1 = -1.77 \times 10^{-6} \MeV ^{-2}, E_1 =  -1.53 \times 10^{-6} \MeV ^{-2}, D_2 = 0, E_2 = 0)$ \footnote{To obtain the reported numerical values from those [$(D_1 = -6.62 , E_1 = -5.74, D_2 = 0,E_2 = 0)$] in Eq.~(29) of~\cite{Xu:2017tsr}, I have used the normalization according to Eq.~(\ref{rel1}),
  i.e. $ (D_i , E_i) \rightarrow [ D_i / (m_{D^*} m_D ), D_i / (m_{D^*} m_D ) ] $.}.
When employed in the present approach, this set of parameters does not yield a bound state for the system $ | [D D^{*}] _{1}^{0}\rangle $, interpreted as the $T_{cc}^+ $ state, neither for other systems, since it is not within the region shown in Fig.~\ref{FIGX1} that allows $S$-wave bound states. 
It is also noteworthy to remark that the findings of~\cite{Xu:2017tsr} obtained with the inclusion of the finite-range potential $V(r)$ due to pion-exchange contributions predicted a bound state with binding energy around 17.5 MeV for the $I=0$ channel, which is clearly greater than that estimated for the  $T_{cc}^+ $; whereas no bound solution is obtained for the  $I=1$ channel. 

Besides, attention should also be deserved to the outcomes from Ref.~\cite{Li:2012ss}, which has utilized one-boson-exchange potential model based on three-body Lagrangians under heavy quark symmetry and $SU(3)$-flavor symmetry, differently from the framework used here. The bound-state solution for the  $ |[D D^*]_{1}^{0}\rangle  $ state is strongly dependent of the cut-off parameter $\Lambda$: for the range $\Lambda = 1.05 - 1.20$ GeV the binding energy varies from 1.24 to 20.98 MeV. Due to the different framework and hypotheses this analysis predicts distinct findings for sectors and channels with respect to those in the present work.

Let me now turn to the case of very recent works which employed the observables of  $T_{cc}^+ $ as input. For example, in the remarkable Refs.~\cite{Albaladejo:2021vln,Du:2021zzh,Ke:2021rxd} can be found coherent results for the  $T_{cc}^+ $ binding energies as in the present study. It is relevant to mention the respective frameworks employed: Ref.~\cite{Albaladejo:2021vln} made use of arguments of heavy quark spin symmetry with four-body interactions, but fitting the input parameters (cut-off, ...) to the $DD \pi$ event distribution from LHCb data; Ref.~\cite{Du:2021zzh} used arguments of heavy quark spin symmetry as well, with the parameters of the interaction being fixed by the observed line shape in the three-body $D^0D^0 \pi^+$ channel; and Ref.~\cite{Ke:2021rxd} worked within the context of Bethe-Salpeter framework. 

On the other hand, there are distinct predictions for other channels $ |[D^{(\ast)} D^{(\ast)}]_{S}^{I}\rangle $ in these recent analyses. 
For example, in the case of the system $ |[D^* D^*]_{1}^{0}\rangle  $, Refs.~\cite{Dai:2021vgf,Albaladejo:2021vln,Du:2021zzh}  found binding energies around 0.5 MeV – 3.2 MeV, which are close to those in Table~\ref{table2} for the sets $(i),(ii)$. It is worth mentioning that Ref.~\cite{Dai:2021vgf} employs an unitarized formalism based on an extension of the local hidden-gauge approach to the charm sector. Also, the analyses in  Refs.~\cite{Dai:2021vgf,Albaladejo:2021vln,Du:2021zzh} have  been restricted to the systems $ |[D^* D^{(*)}]_{S}^{I}\rangle  $; so no comparison has been made with them for other other channels and sectors. In contrast, Ref.~\cite{Ke:2021rxd} reports no bound solution for  $ |[D^* D^*]_{1}^{0}\rangle  $, differently of our findings. Besides, the system $ | [D^* D^{*}] _{0}^{1}\rangle $ acquires in~\cite{Ke:2021rxd} a binding energy of almost 5 times greater than that obtained with the set  $ (i) $ in Table~\ref{table2}.

In the end, the observation of the first doubly-charmed tetraquark state $T_{cc}^+ $ has naturally opened a new wide window of opportunities for the study of the distinct families of doubly heavy tetraquark states. Before, they appeared just as speculations, but now we have some important observables (like mass and width) of the $T_{cc}^+ $  which could provide us some insight on the nature of these states. In this sense, since these observables are compatible with the interpretation of $T_{cc}^+ $ as a shallow-bound $ D D^{*} $ state, they became a benchmark to determine the LECs and free parameters, and preceding studies like~\cite{Xu:2017tsr} should be revisited in order to make their predictions more precise and compatible with the cited observables.  In addition, the existence of a shallow-bound $ D D^{*} $  state naturally encourages the use of the same scenario to analyze and predict the existence of possible loosely-bound states in other channels and sectors. The present subsection and next ones are thus intended to make a contribution on this subject.

\subsection{ $ |[\bar{B}^{(*)} \bar{B}^{(*)}]_{S}^{I}\rangle $ systems}

\begin{figure}[!ht]
	\centering
           \includegraphics[{width=8.0cm}]{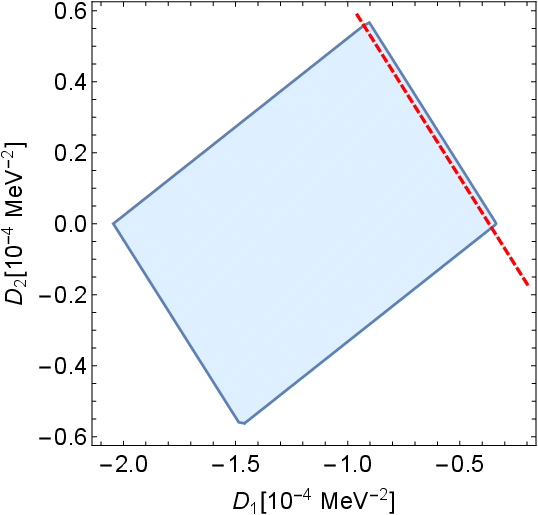}
\caption{ The same as in Fig.~\ref{FIGX1}, but for  $ |[\bar{B}^{(*)} \bar{B}^{(*)}]_{S}^{I}\rangle $ systems. The dashed line represents the relation between the parameters which reproduces the quantities used for the system $ | [\bar{B} \bar{B}^{*}] _{1}^{0}\rangle $, interpreted as the $T_{bb}^+ $ state.
}
	\label{FIGX2}
\end{figure}

The analysis for doubly-bottomed systems can be performed similarly to that for the charmed ones. But due to the difference of the mass spectrum between the $ |[D^{(*)} D^{(*)}]_{S}^{I}\rangle $ and $ |[\bar{B}^{(*)} \bar{B}^{(*)}]_{S}^{I}\rangle $, 
the renormalization procedure must be carried out separately in each sector. Then, assuming the same assumptions as in previous situation, in Fig.~\ref{FIGX2} is displayed the light shaded area specifying the intersection region in $(D_1,D_2)$-parameter space (with $E_1, E_2 = 0$) for allowed solutions of shallow $S$-wave bound states according to  Eq.~(\ref{allstates}). We remark that this region is reduced with respect to that in Fig.~\ref{FIGX1}, meaning that the couplings should acquire smaller values to have doubly-bottomed bound states.

Despite the theoretical predictions for this sector~(see e.g. \cite{Wang:2018atz,Barnes:1999hs,Ebert:2007rn,Zhang:2007mu,Vijande:2009kj,Yang:2009zzp,Du:2012wp,Luo:2017eub,Eichten:2017ffp,Liu:2019zoy,Yu:2019sxx,Dai:2022ulk}), in the experimental landscape there is no any doubly-bottomed bound state observed yet. In the lack of data to fix or at least to constraint the parameter space, we follow as a guide the possible existence of the $T_{bb}^+ $ state, which would be the analogous structure of the $T_{cc}^+ $ in the doubly-bottomed sector with quantum numbers $I(J^P) =  0(1^+)$. In view of this, working with the prediction of a possible loosely-bound state for the $ | [\bar{B} \bar{B}^{*}] _{1}^{0}\rangle $ system, with a binding energy $E_b = 3.1$ MeV as in Ref.~\cite{Yu:2019sxx} (in consonance with our limit of validity), and proceeding as previously, it is straightforward to get $V^{[\bar{B} \bar{B}^{*}] _{1}^{0}} = 7.39 \times 10^{-5} \MeV ^{-2}$. 
In the $(D_1,D_2)$-parameter space with $E_1, E_2 = 0$, this relation is represented in Fig.~\ref{FIGX2} by the dashed line.

As earlier, using the parameters to reproduce the prediction for the $ | [\bar{B} \bar{B}^{*}] _{1}^{0}\rangle $ bound structure as basis, then Table~\ref{table3} summarizes the results obtained for the mass, binding energy and scattering length of the 
$ |[\bar{B}^{(*)} \bar{B}^{(*)}]_{S}^{I}\rangle $ with three sets of parameters in a such way that the first two ones have their magnitudes along the dashed line inside the region of intersection in Fig.~\ref{FIGX2}, whereas the latter one has its magnitude along the line but outside this area.
As expected, all the bound states are just near the thresholds of the related channels, depending on the set of parameters. These findings are in agreement with the predictions for the $ | [\bar{B} \bar{B}^{*}] _{1}^{0}\rangle $ and not far for the $ | [\bar{B}^{*} \bar{B}^{*}] _{1}^{0}\rangle $ from Ref.~\cite{Yu:2019sxx}, which makes use of interchange constituent model. On the other hand, the mentioned Ref. found near-threshold virtual solutions for the other states. Besides, within the one-boson-exchange potential model, Ref.~\cite{Li:2012ss} obtained (for smaller values of the cut-off) results near those of the set $(i)$. In contrast, other approaches reported for example in Refs.~\cite{Ke:2021rxd,Dai:2022ulk} estimated larger binding energies.

\begin{center}
\begin{table}[h!]
 \caption{The same as in Table~\ref{table2}, but for  Relevant quantities for the $ |[\bar{B}^{(*)} \bar{B}^{(*)}]_{S}^{I}\rangle $ systems. The results are obtained in the context of three different set of parameters $(D_1,E_1,D_2,E_2)$ (values in $ \MeV ^{-2}$): $(i) (-0.0001, 0, 0.00000302795, 0);$ $ (ii) (-0.0001, -0.000021, 0, 0) ;$ and $ (iii) (-0.0006, 0.00059, 0.0006, 0.0006) $. Bold column indicates the state interpreted as the tetraquark $T_{bb}^+ $, which has been used as basis. } 
\begin{tabular}{M{60pt}|M{35pt}|M{40pt}|M{35pt}|M{35pt}}
\hline\hline \vspace{3pt}
System    & Set & $ M $ &  $ E_b $  & $a_s$  \\[3pt]
   ($M_{Th}$)  &  & &   &  \\[3pt]
\hline 
\hline \vspace{3pt}
$ |[\bar{B} \bar{B}]_{0}^{1}\rangle  $  & (i) & 10556.12 &  2.68 &  1.66  \\[2pt]
 (10558.80) &   (ii)  & 10558.51  & 0.29  &  5.03
 \\[2pt]
  &   (iii)  & -  & -  &  -
 \\[2pt]\hline \vspace{3pt}
$\mathbf{ |[\bar{B} \bar{B}^*]_{1}^{0}\rangle  }$  & \textbf{(i)} & \textbf{10601.50} & \textbf{3.10} & \textbf{1.54}  \\[3pt]
\textbf{(10604.60)}  &   \textbf{(ii)} & \textbf{10601.50}  & \textbf{3.10}  & \textbf{1.54} 
 \\[2pt]
  &    \textbf{(iii)}    & \textbf{10601.50}  & \textbf{3.10}  & \textbf{1.54}
 \\[2pt] \hline \vspace{3pt}
 $ |[\bar{B} \bar{B}^*]_{1}^{1}\rangle  $  & (i) & 10602.31 &  2.29  & 1.79   \\[2pt]
(10604.60)  &   (ii) & 10604.31 &  0.29  & 5.03 
 \\[2pt] 
 &   (iii)  & -  & -  &  -
 \\[2pt] \hline \vspace{3pt}
$ |[\bar{B}^* \bar{B}^*]_{0}^{1}\rangle  $  & (i) & 10648.43 & 1.97 &  1.92 \\[2pt]
(10650.40)  &   (ii)  & 10650.11  & 0.29  & 5.05 
 \\[2pt]
  &   (iii)  & -  & -  &  -
 \\[2pt]\hline \vspace{3pt}
$ |[\bar{B}^* \bar{B}^*]_{1}^{0}\rangle  $  &  (i) & 10648.14 &  2.26 & 1.80  \\[2pt]
(10650.40)   &   (ii)  & 10647.34  &  3.06 & 1.54  
\\[2pt]&   (iii)  & -  & -  &  -
 \\[2pt]
 \hline \vspace{3pt}
$ |[\bar{B}^* \bar{B}^*]_{2}^{1}\rangle  $    &   (i)  & 10647.34  & 3.06 & 1.54 \\[2pt]
 (10650.40) & (ii)  & 10650.11  & 0.29  & 11.08 
  \\[3pt]
 &   (iii)  & -  & -  &  -
 \\[2pt]\hline 
\hline \end{tabular}
  \label{table3}
\end{table}
\end{center}

\subsection{ $ |[D^{(*)} \bar{B}^{(*)}]_{S}^{I}\rangle $ systems}

\begin{figure}[!ht]
	\centering
           \includegraphics[{width=8.0cm}]{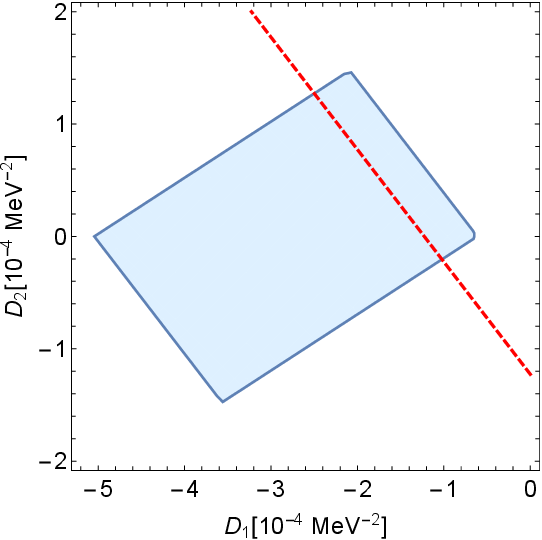}
\caption{ The same as in Fig.~\ref{FIGX1}, but for  $ |[D^{(*)} \bar{B}^{(*)}]_{S}^{I}\rangle $ systems. The dashed line represents the relation between the parameters which reproduces the values used for the system $ | [\bar{B} \bar{B}^{*}] _{1}^{0}\rangle $, interpreted as the $T_{cb}^+ $ state (the equivalent structure in the doubly-bottomed sector to the $T_{cc}^+ $).
}
	\label{FIGX3}
\end{figure}

This study is finished  with the $ |[D^{(*)} \bar{B}^{(*)}]_{S}^{I}\rangle $ systems. However, differently from the preceding cases, the di-meson states with both open charm and bottom do not have the restrictions related to the selection rule mentioned earlier. In other words, all the systems are allowed. Then, proceeding with the renormalization scheme within this sector, and assuming one more time the assumptions as in previous situations, we get the light shaded area shown in Fig.~\ref{FIGX3} representing the intersection region in $(D_1,D_2)$-parameter space (with $E_1, E_2 = 0$) for solutions of shallow $S$-wave bound states according to  Eqs.~(\ref{iso}) and~(\ref{tri}). 
It can be inferred that the boundaries of this region give potentials with intermediate magnitudes among those of the sectors studied before.

Again, due to the lack of data to fix or at least to constraint the parameter space, it is considered as a guide the predictions from Ref.~\cite{Li:2012ss,Sakai:2017avl} of a possible loosely-bound state for the $ | [D \bar{B}^{*}] _{1}^{0}\rangle $ with binding energy in the range $1.21 - 6.30$ MeV. Using $E_b = 2.0$ MeV, and proceeding as previously, it is simple to get $V^{[D \bar{B}^{*}] _{1}^{0}} = 2.44 \times 10^{-4} \MeV ^{-2}$. 
In the $(D_1,D_2)$-parameter space with $E_1, E_2 = 0$, this relation is represented in Fig.~\ref{FIGX2} by the dashed line.

As earlier, with the parameters reproducing the prediction for the $ | [D \bar{B}^{*}] _{1}^{0}\rangle $ bound structure as basis, then Table~\ref{table4} summarizes the results obtained for the mass, binding energy and scattering length of the 
$ |[D^{(*)} \bar{B}^{(*)}]_{S}^{I}\rangle $ with three sets of parameters in a such way that the first two ones have their magnitudes along the dashed line inside the region of intersection in Fig.~\ref{FIGX2}, whereas the latter one has its magnitude along the line but outside this area. According to the set of parameters,  bound states solutions just near the thresholds are or not obtained.

\begin{center}
\begin{table}[h!]
 \caption{The same as in Table~\ref{table2}, but for  Relevant quantities for the $ |[D^{(*)} \bar{B}^{(*)}]_{S}^{I}\rangle $ systems. The results are obtained in the context of three different set of parameters $(D_1,E_1,D_2,E_2)$ (values in $ \MeV ^{-2}$): $(i) (-0.0002, 0, 0.000078, 0);$ $ (ii) (-0.0002, -0.0000259, 0, 0);$ and $ (iii) (-0.001, -0.000959, 0.001, 0.001) $. Bold column indicates the state interpreted as the tetraquark $T_{bc}^+ $, which has been used as basis. } 
\begin{tabular}{M{60pt}|M{35pt}|M{35pt}|M{35pt}|M{35pt}}
\hline\hline \vspace{3pt}
System    & Set & $ M $ &  $ E_b $  & $a_s$  \\[3pt]
   ($M_{Th}$)  &  & &   &  \\[3pt]
\hline 
\hline \vspace{3pt}
$ |[D \bar{B}]_{0}^{0}\rangle  $  & (i) & 7145.91 &  0.75 &  4.32 \\[2pt]
 (7146.66) &   (ii)  & 7144.65  & 2.01  &  2.64
 \\[2pt]
  &   (iii)  & -  & -  &  -
 \\[2pt]\hline \vspace{3pt}
$ |[D \bar{B}]_{0}^{1}\rangle  $  & (i) & 7145.91 &  0.75 &  4.32  \\[2pt]
 (7146.66) &   (ii)  & 7146.07  & 0.59  &  4.89
 \\[2pt]
  &   (iii)  & -  & -  &  -
 \\[2pt]\hline \vspace{3pt}
$\mathbf{ |[D \bar{B}^*]_{1}^{0}\rangle  }$  & \textbf{(i)} & \textbf{7190.44} & \textbf{2.00} & \textbf{2.65}  \\[3pt]
\textbf{7192.44}  &   \textbf{(ii)} & \textbf{7190.44}  & \textbf{2.00} & \textbf{2.65}
 \\[2pt]
  &    \textbf{(iii)}    & \textbf{7190.44}  & \textbf{2.00} & \textbf{2.65}
 \\[2pt] \hline \vspace{3pt}
$ |[D \bar{B}^*]_{1}^{1}\rangle  $  &  (i) & 7192.05 & 0.39 & 6.02  \\[3pt]
(7192.44)  &   (ii) & 7191.85 & 0.59  & 4.90  
 \\[2pt]
  &     (iii)    &  - & -  & -
 \\[2pt] \hline \vspace{3pt}
 $ |[D^* \bar{B}]_{1}^{0}\rangle  $  & (i) & 7286.26 &  1.72  & 2.79   \\[2pt]
(7287.98)  &   (ii) &  7286.26 &  1.72  & 2.79
 \\[2pt] 
 &   (iii)  &  7286.26  & 1.72  &  2.79
  \\[2pt] \hline \vspace{3pt}
 $ |[D^* \bar{B}]_{1}^{1}\rangle  $  & (i) & 7287.65 &  0.33  & 6.34   \\[2pt]
(7287.98)  &   (ii) & 7287.48 &  0.50  & 5.15 
 \\[2pt] 
 &   (iii)  & -  & -  &  -
  \\[2pt] \hline \vspace{3pt}
$ |[D^* \bar{B}^*]_{0}^{0}\rangle  $  & (i) & 7333.56 & 0.20 &  8.13 \\[2pt]
(7333.76)  &   (ii)  & 7332.06  & 1.70  & 2.79 
 \\[2pt]
  &   (iii)  & -  & -  &  -
 \\[2pt]\hline \vspace{3pt}
$ |[D^* \bar{B}^*]_{0}^{1}\rangle  $  & (i) & 7333.56 & 0.20 &  8.13 \\[2pt]
(7333.76)  &   (ii)  & 7333.26  & 0.50  & 5.17 
 \\[2pt]
  &   (iii)  & -  & -  &  -
 \\[2pt]\hline \vspace{3pt}
$ |[D^* \bar{B}^*]_{1}^{0}\rangle  $  &  (i) & 7333.43 &  0.33 & 6.35  \\[2pt]
(7333.76)   &   (ii)  & 7332.06  &  1.70 & 2.79  
\\[2pt]&   (iii)  & -  & -  &  -
 \\[2pt]
 \hline \vspace{3pt}
$ |[D^* \bar{B}^*]_{1}^{1}\rangle  $  &  (i) & 7333.43 &  0.33 & 6.35  \\[2pt]
(7333.76)   &   (ii)  & 7333.26  &  0.50 & 5.17  
\\[2pt]&   (iii)  & -  & -  &  -
 \\[2pt]
 \hline \vspace{3pt}
$ |[D^* \bar{B}^*]_{2}^{0}\rangle  $    &   (i)  & 7332.06  & 1.70 & 2.79 \\[2pt]
 (7333.76) & (ii)  & 7332.06  & 1.70  & 2.79 
  \\[3pt]
 &   (iii)  & -  & 1.70  &  2.79
  \\[2pt]
\hline \vspace{3pt}
$ |[D^* \bar{B}^*]_{2}^{1}\rangle  $    &   (i)  & 7332.06  &  1.70 & 1.54 \\[2pt]
 (7333.76) & (ii)  & 7333.26  & 0.50  & 5.17 
  \\[3pt]
 &   (iii)  & -  & -  &  -
 \\[2pt]\hline 
\hline \end{tabular}
  \label{table4}
\end{table}
\end{center}

\section{Discussion and Concluding Remarks}
\lb{Conclusions}

Summarizing, in this work the Heavy-Meson Effective Theory has been used to perform a simplified field-theoretical study of possible deuteron-like  $D^{(\ast)} D ^{(\ast)}, \bar{B}^{(\ast)} \bar{B}^{(\ast)} $ and $D^{(\ast)} \bar{B}^{(\ast)} $ molecules. The main motivation came from the recently observed tetraquark $T_{cc}^+ $ state, whose observables have been employed as input to fix the potential associated to the shallow-bound $ D D^{*} $ system with $I(J^P) =  0(1^+)$. The conditions for the formation of other loosely-bound states have been analyzed: in particular,  estimations of their masses, binding energies and scattering lengths have been 
performed as functions of interaction strength in some specific renormalization 
schemes. The validity of the present approach is within the context of contact-range interactions, in which the pion-exchange contribution is not relevant and bound states acquire larger scattering lengths.

In the doubly-charmed sector, the findings obtained suggest that the existence and the properties of bound states just near the thresholds are strongly dependent on the set of parameters. The only use of the information of the $T_{cc}^+ $ as a $ | [D D^{*}] _{1}^{0}\rangle $ system is insufficient to fix the four couplings present in the formalism, which predicts other possible $ |[D^{(*)} D^{(*)}]_{S}^{I}\rangle $ bound structures or not according to their magnitudes. Notwithstanding, results engendered by specific sets of parameters have been compatible with other predictions available in literature within other frameworks. In this sense, more experimental data are needed to determine if more $ |[D^{(*)} D^{(*)}]_{S}^{I}\rangle $ systems can be interpreted as bound states, and if so what are their characteristics and if they (or some of them) can fit into this simplified model of contact-range interaction. This effective approach might be in principle useful in the context of situations where the interest is mostly on the qualitative and general aspects of these structures. This discussion obviously applies for the $ \bar{B}^{(\ast)} \bar{B}^{(\ast)} $ and $D^{(\ast)} \bar{B}^{(\ast)} $ sectors.

%

\begin{center}
{\bf ACKNOWLEDGMENTS}
\end{center}

 I would like to thank the Brazilian funding agencies for their
  financial support: CNPq (LMA: contracts 309950/2020-1 and 400546/2016-7) and FAPESB (LMA: contract INT0007/2016).  

I thank Eulogio Oset for reading the draft and pointing out improvements.



\begin{thebibliography}{9} \label{Bibliography}
\bibitem{LHCb:2021vvq}   R.~Aaij \textit{et al.} [LHCb],
                         [arXiv:2109.01038 [hep-ex]].

\bibitem{LHCb:2021auc}   R.~Aaij \textit{et al.} [LHCb],
                         [arXiv:2109.01056 [hep-ex]].


\bibitem{Gelman:2002wf}  B.~A.~Gelman and S.~Nussinov,
                         Phys. Lett. B \textbf{551}, 296 (2003).



\bibitem{Janc:2004qn}    D.~Janc and M.~Rosina,
                         Few Body Syst. \textbf{35}, 175 (2004). 

\bibitem{Vijande:2003ki} J.~Vijande, F.~Fernandez, A.~Valcarce and
                         B.~Silvestre-Brac,
                         Eur. Phys. J. A \textbf{19}, 383 (2004). 


\bibitem{Navarra:2007yw} F.~S.~Navarra, M.~Nielsen and S.~H.~Lee,
                         Phys. Lett. B \textbf{649}, 166 (2007). 

\bibitem{Vijande:2007rf} J.~Vijande, E.~Weissman, A.~Valcarce and N.~Barnea,
                         Phys. Rev. D \textbf{76}, 094027 (2007). 

\bibitem{Ebert:2007rn}   D.~Ebert, R.~N.~Faustov, V.~O.~Galkin and W.~Lucha,
                         Phys. Rev. D \textbf{76}, 114015 (2007). 

\bibitem{Lee:2009rt}     S.~H.~Lee and S.~Yasui,
                         Eur. Phys. J. C \textbf{64}, 283 (2009).


\bibitem{Yang:2009zzp}     Y.~Yang, C.~Deng, J.~Ping and T.~Goldman,
                           Phys. Rev. D \textbf{80}, 114023 (2009). 

\bibitem{Hong:2018mpk}     J.~Hong, S.~Cho, T.~Song and S.~H.~Lee,
                           Phys. Rev. C \textbf{98},  014913 (2018).

\bibitem{Hudspith:2020tdf} R.~J.~Hudspith, B.~Colquhoun, A.~Francis,
                           R.~Lewis and K.~Maltman,
                           Phys. Rev. D \textbf{102}, 114506 (2020). 

\bibitem{Cheng:2020wxa}    J.~B.~Cheng, S.~Y.~Li, Y.~R.~Liu, Z.~G.~Si
                           and T.~Yao,
                           Chin. Phys. C \textbf{45}, 043102 (2021). 

\bibitem{Qin:2020zlg}      Q.~Qin, Y.~F.~Shen and F.~S.~Yu,
                           Chin. Phys. C \textbf{45}, 103106 (2021). 

\bibitem{Agaev:2021vur}    S.~S.~Agaev, K.~Azizi and H.~Sundu,
                           [arXiv:2108.00188 [hep-ph]].


\bibitem{Dong:2021bvy}     X.~K.~Dong, F.~K.~Guo and B.~S.~Zou,
                           [arXiv:2108.02673 [hep-ph]].

\bibitem{Feijoo:2021ppq}
A.~Feijoo, W.~H.~Liang and E.~Oset,
Phys. Rev. D \textbf{104} (2021) no.11, 114015
doi:10.1103/PhysRevD.104.114015
[arXiv:2108.02730 [hep-ph]].

\bibitem{Huang:2021urd}
Y.~Huang, H.~Q.~Zhu, L.~S.~Geng and R.~Wang,
[arXiv:2108.13028 [hep-ph]].


\bibitem{Li:2021zbw} N.~Li, Z.~F.~Sun, X.~Liu and S.~L.~Zhu,
                     Chin. Phys. Lett. \textbf{38}, 092001 (2021). 


\bibitem{Ren:2021dsi}
H.~Ren, F.~Wu and R.~Zhu,
Adv. High Energy Phys. \textbf{2022} (2022), 9103031
doi:10.1155/2022/9103031
[arXiv:2109.02531 [hep-ph]].

\bibitem{Xin:2021wcr}
Q.~Xin and Z.~G.~Wang,
[arXiv:2108.12597 [hep-ph]].


\bibitem{Yang:2021zhe}
G.~Yang, J.~Ping and J.~Segovia,
Phys. Rev. D \textbf{104} (2021) no.9, 094035
doi:10.1103/PhysRevD.104.094035
[arXiv:2109.04311 [hep-ph]].


\bibitem{Meng:2021jnw} L.~Meng, G.~J.~Wang, B.~Wang and S.~L.~Zhu,
                       Phys. Rev. D \textbf{104}, 051502 (2021). 

\bibitem{Ling:2021bir}
X.~Z.~Ling, M.~Z.~Liu, L.~S.~Geng, E.~Wang and J.~J.~Xie,
Phys. Lett. B \textbf{826} (2022), 136897
doi:10.1016/j.physletb.2022.136897
[arXiv:2108.00947 [hep-ph]].

\bibitem{Fleming:2021wmk}
S.~Fleming, R.~Hodges and T.~Mehen,
Phys. Rev. D \textbf{104} (2021) no.11, 116010
doi:10.1103/PhysRevD.104.116010
[arXiv:2109.02188 [hep-ph]].

\bibitem{Jin:2021cxj}
Y.~Jin, S.~Y.~Li, Y.~R.~Liu, Q.~Qin, Z.~G.~Si and F.~S.~Yu,
Phys. Rev. D \textbf{104} (2021) no.11, 114009
doi:10.1103/PhysRevD.104.114009
[arXiv:2109.05678 [hep-ph]].

\bibitem{Azizi:2021aib}
K.~Azizi and U.~\"Ozdem,
Phys. Rev. D \textbf{104} (2021) no.11, 114002
doi:10.1103/PhysRevD.104.114002
[arXiv:2109.02390 [hep-ph]].


\bibitem{Hu:2021gdg}
Y.~Hu, J.~Liao, E.~Wang, Q.~Wang, H.~Xing and H.~Zhang,
[arXiv:2109.07733 [hep-ph]].


\bibitem{Abreu:2021jwm}
L.~M.~Abreu, F.~S.~Navarra, M.~Nielsen and H.~P.~L.~Vieira,
Eur. Phys. J. C \textbf{82} (2022) no.4, 296
doi:10.1140/epjc/s10052-022-10238-8
[arXiv:2110.11145 [hep-ph]].

\bibitem{Abreu:2022lfy}
L.~M.~Abreu, F.~S.~Navarra and H.~P.~L.~Vieira,
[arXiv:2202.10882 [hep-ph]].

\bibitem{Dai:2021vgf}
L.~R.~Dai, R.~Molina and E.~Oset,
Phys. Rev. D \textbf{105} (2022) no.1, 016029
doi:10.1103/PhysRevD.105.016029
[arXiv:2110.15270 [hep-ph]].


\bibitem{Albaladejo:2021vln}
M.~Albaladejo,
Phys. Lett. B \textbf{829} (2022), 137052
doi:10.1016/j.physletb.2022.137052
[arXiv:2110.02944 [hep-ph]].

\bibitem{Du:2021zzh}
M.~L.~Du, V.~Baru, X.~K.~Dong, A.~Filin, F.~K.~Guo, C.~Hanhart, A.~Nefediev, J.~Nieves and Q.~Wang,
Phys. Rev. D \textbf{105} (2022) no.1, 014024
doi:10.1103/PhysRevD.105.014024
[arXiv:2110.13765 [hep-ph]].










%
%

\bibitem{Voloshin} M.B. Voloshin and L. B. Okun, JETP Lett. {23} (1976) 333.

\bibitem{Tornqvist1} N.A. Tornqvist, Nuovo Cim. A {107} (1994) 2471, arXiv:hep-ph
/9310225 [hep-ph].

\bibitem{Tornqvist2} N.A. Tornqvist, Z. Phys. C {61} (1994)  525, arXiv:hep-ph
/9310247.

\bibitem{Tornqvist3} N. A. Tornqvist, Phys. Lett. B {590} (2004) 209,
arXiv:hep-ph/0402237.


\bibitem{AlFiky} M.T. Alfiky, F. Gabbiani and A. A. Petrov, 
Phys. Lett. B {640} (2006) 238, arXiv:hep-ph/0506141. 

\bibitem{Braaten1}  E. Braaten and M. Lu, Phys. Rev. D {76} (2007) 094028,
arXiv:0709.2697 [hep-ph].

\bibitem{Dong1} Y. b. Dong, A. Faessler, T. Gutsche, and
V. E. Lyubovitskij, Phys. Rev. D {77} (2008) 094013,
arXiv:0802.3610 [hep-ph]. 

%
%
\bibitem{Braaten2} E. Braaten and J. Stapleton, Phys. Rev. D {81} (2010)
014019, arXiv:0907.3167 [hep-ph].

\bibitem{Nieves} J. Nieves and M.P. Valderrama, Phys. Rev. D {86} (2012) 056004, 
arXiv:1204.2790. 

\bibitem{Hidalgo} C. Hidalgo-Duque, J. Nieves, M. Pavon Valderrama, 
Phys. Rev. D {87} (2013) 076006, 	arXiv:1210.5431 [hep-ph].

\bibitem{Guo} F.-K Guo, C. Hidalgo-Duque, J. Nieves, M. P. Valderrama, Phys. Rev. 
D {88} (2013) 054007, arXiv:1303.6608 [hep-ph].

\bibitem{Alberto} A. Martinez Torres, K. P. Khemchandani, F. S. Navarra, M. Nielsen, 
L. M. Abreu, Phys. Rev. D {90} (2014) 114023, arXiv:1405.7583 [hep-ph].

\bibitem{XProd2}     L. M. Abreu, K. P. Khemchandani, A. Martinez Torres, 
                     F. S. Navarra and M. Nielsen, 
                     Phys. Lett. B {\bf 761}, 303 (2016).

\bibitem{Braaten:2020iqw}
E.~Braaten, L.~P.~He, K.~Ingles and J.~Jiang,
Phys. Rev. D \textbf{103} (2021) no.7, L071901
doi:10.1103/PhysRevD.103.L071901
[arXiv:2012.13499 [hep-ph]].

\bibitem{Guo:2017jvc}
F.~K.~Guo, C.~Hanhart, U.~G.~Mei\ss{}ner, Q.~Wang, Q.~Zhao and B.~S.~Zou,
Rev. Mod. Phys. \textbf{90} (2018) no.1, 015004
doi:10.1103/RevModPhys.90.015004
[arXiv:1705.00141 [hep-ph]].

\bibitem{Brambilla:2019esw}
N.~Brambilla, S.~Eidelman, C.~Hanhart, A.~Nefediev, C.~P.~Shen, C.~E.~Thomas, A.~Vairo and C.~Z.~Yuan,
Phys. Rept. \textbf{873} (2020), 1-154
doi:10.1016/j.physrep.2020.05.001
[arXiv:1907.07583 [hep-ex]].

%
%
%
%
%
%
%
%
%
%


\bibitem{Zhang:2007mu}
M.~Zhang, H.~X.~Zhang and Z.~Y.~Zhang,
Commun. Theor. Phys. \textbf{50} (2008), 437-440
doi:10.1088/0253-6102/50/2/31
[arXiv:0711.1029 [nucl-th]].


\bibitem{Molina:2010tx}
R.~Molina, T.~Branz and E.~Oset,
Phys. Rev. D \textbf{82} (2010), 014010
doi:10.1103/PhysRevD.82.014010
[arXiv:1005.0335 [hep-ph]].



\bibitem{Li:2012ss}
N.~Li, Z.~F.~Sun, X.~Liu and S.~L.~Zhu,
Phys. Rev. D \textbf{88} (2013) no.11, 114008
doi:10.1103/PhysRevD.88.114008
[arXiv:1211.5007 [hep-ph]].


\bibitem{Du:2012wp}
M.~L.~Du, W.~Chen, X.~L.~Chen and S.~L.~Zhu,
Phys. Rev. D \textbf{87} (2013) no.1, 014003
doi:10.1103/PhysRevD.87.014003
[arXiv:1209.5134 [hep-ph]].

\bibitem{Luo:2017eub}
S.~Q.~Luo, K.~Chen, X.~Liu, Y.~R.~Liu and S.~L.~Zhu,
Eur. Phys. J. C \textbf{77} (2017) no.10, 709
doi:10.1140/epjc/s10052-017-5297-4
[arXiv:1707.01180 [hep-ph]].

\bibitem{Sakai:2017avl}
S.~Sakai, L.~Roca and E.~Oset,
Phys. Rev. D \textbf{96} (2017) no.5, 054023
doi:10.1103/PhysRevD.96.054023
[arXiv:1704.02196 [hep-ph]].

\bibitem{Eichten:2017ffp}
E.~J.~Eichten and C.~Quigg,
Phys. Rev. Lett. \textbf{119} (2017) no.20, 202002
doi:10.1103/PhysRevLett.119.202002
[arXiv:1707.09575 [hep-ph]].


\bibitem{Xu:2017tsr}
H.~Xu, B.~Wang, Z.~W.~Liu and X.~Liu,
Phys. Rev. D \textbf{99} (2019) no.1, 014027
[erratum: Phys. Rev. D \textbf{104} (2021) no.11, 119903]
doi:10.1103/PhysRevD.99.014027
[arXiv:1708.06918 [hep-ph]].

\bibitem{Wang:2018atz}
B.~Wang, Z.~W.~Liu and X.~Liu,
Phys. Rev. D \textbf{99} (2019) no.3, 036007
doi:10.1103/PhysRevD.99.036007
[arXiv:1812.04457 [hep-ph]].


\bibitem{Liu:2019zoy}
Y.~R.~Liu, H.~X.~Chen, W.~Chen, X.~Liu and S.~L.~Zhu,
Prog. Part. Nucl. Phys. \textbf{107} (2019), 237-320
doi:10.1016/j.ppnp.2019.04.003
[arXiv:1903.11976 [hep-ph]].

\bibitem{Barnes:1999hs}
T.~Barnes, N.~Black, D.~J.~Dean and E.~S.~Swanson,
Phys. Rev. C \textbf{60} (1999), 045202
doi:10.1103/PhysRevC.60.045202
[arXiv:nucl-th/9902068 [nucl-th]].

\bibitem{Vijande:2009kj}
J.~Vijande, A.~Valcarce and N.~Barnea,
Phys. Rev. D \textbf{79} (2009), 074010
doi:10.1103/PhysRevD.79.074010
[arXiv:0903.2949 [hep-ph]].

\bibitem{Yu:2019sxx}
M.~T.~Yu, Z.~Y.~Zhou, D.~Y.~Chen and Z.~Xiao,
Phys. Rev. D \textbf{101} (2020) no.7, 074027
doi:10.1103/PhysRevD.101.074027
[arXiv:1912.07348 [hep-ph]].

\bibitem{Chen:2021cfl}
K.~Chen, R.~Chen, L.~Meng, B.~Wang and S.~L.~Zhu,
[arXiv:2109.13057 [hep-ph]].

\bibitem{Zhao:2021cvg}
M.~J.~Zhao, Z.~Y.~Wang, C.~Wang and X.~H.~Guo,
Phys. Rev. D \textbf{105} (2022) no.9, 096016
doi:10.1103/PhysRevD.105.096016
[arXiv:2112.12633 [hep-ph]].

\bibitem{Dai:2022ulk}
L.~R.~Dai, E.~Oset, A.~Feijoo, R.~Molina, L.~Roca, A.~M.~Torres and K.~P.~Khemchandani,
Phys. Rev. D \textbf{105} (2022) no.7, 074017
doi:10.1103/PhysRevD.105.074017
[arXiv:2201.04840 [hep-ph]].



\bibitem{Manohar} A.V. Manohar, M.B. Wise, {\it  Heavy quark physics },
 Cambridge Monographs on Particle Physics, Nuclear Physics, and Cosmology (Cambridge
 University Press, Cambridge, 2000).    
 
\bibitem{Valderrama} M.P. Valderrama, Phys.Rev. D {85} (2012) 114037, arXiv:1204.2400. 


\bibitem{Abreu:2015jma}
L.~M.~Abreu,
Nucl. Phys. A \textbf{940} (2015), 1-20
doi:10.1016/j.nuclphysa.2015.03.015
[arXiv:1504.01801 [hep-ph]].

\bibitem{Abreu:2016dfe} 
  L.~M.~Abreu,
  PTEP {\bf 2016} (2016)  103B01
  [arXiv:1608.08165 [hep-ph]].

\bibitem{Abreu:2016xlr} 
  L.~M.~Abreu and A.~Lafayette Vasconcellos,  Phys.\ Rev.\ D {\bf 94} (2016)  096009.
  


\bibitem{Abreu:2017pos}
L.~M.~Abreu, F.~S.~Navarra, M.~Nielsen and A.~L.~Vasconcellos,
Eur. Phys. J. C \textbf{78} (2018) no.9, 752
doi:10.1140/epjc/s10052-018-6182-5
[arXiv:1711.05205 [hep-ph]].
 



\bibitem{Ke:2021rxd}
H.~W.~Ke, X.~H.~Liu and X.~Q.~Li,
Eur. Phys. J. C \textbf{82} (2022) no.2, 144
doi:10.1140/epjc/s10052-022-10092-8
[arXiv:2112.14142 [hep-ph]].



\bibitem{ParticleDataGroup:2020ssz}
P.~A.~Zyla \textit{et al.} [Particle Data Group],
PTEP \textbf{2020} (2020) no.8, 083C01
doi:10.1093/ptep/ptaa104

\bibitem{Fleming:2007rp}
S.~Fleming, M.~Kusunoki, T.~Mehen and U.~van Kolck,
Phys. Rev. D \textbf{76} (2007), 034006
doi:10.1103/PhysRevD.76.034006
[arXiv:hep-ph/0703168 [hep-ph]].


\bibitem{Meng:2022ozq}
L.~Meng, B.~Wang, G.~J.~Wang and S.~L.~Zhu,
[arXiv:2204.08716 [hep-ph]].
                         




%
%
%
%




\end{thebibliography}
\end{document}